\documentclass[twocolumn,aps,superscriptaddress,showpacs,nofootinbib,floatfix]{revtex4}
\usepackage{epsfig,bm,feynmf}
\usepackage{graphics}
\usepackage{amsmath}
\usepackage[normalem]{ulem}  
\usepackage[dvips]{color} 

\renewcommand\sout{\bgroup \color{red} \ULdepth=-.5ex \ULset}

\begin{document}


\title{Quarkonium formation time in relativistic heavy-ion collisions}


\author{Taesoo Song}\email{song@fias.uni-frankfurt.de}
\affiliation{Frankfurt Institute for Advanced Studies and Institute for Theoretical Physics, Johann Wolfgang Goethe Universit\"{a}t, Frankfurt am Main, Germany}
\author{Che Ming Ko}\email{ko@comp.tamu.edu}
\affiliation{Cyclotron Institute and Department of Physics and Astronomy, Texas A$\&$M University, College Station, TX 77843, USA}
\author{Su Houng Lee}\email{suhoung@yonsei.ac.kr}
\affiliation{Institute of Physics and Applied Physics, Yonsei
University, Seoul 120-749, Korea}


\begin{abstract}
We calculate the quarkonium formation time in relativistic heavy-ion collisions from the space-time correlator of heavy quark vector currents in a hydrodynamics background with the initial nonequilibrium stage expanding only in the longitudinal direction. Using in-medium quarkonia properties determined with the heavy quark potential taken to be the free energy from lattice calculations and the fact that quarkonia can only be formed below their dissociation temperatures due to color screening, we find that $\Upsilon$(1S), $\Upsilon$(2S), $\Upsilon$(3S), $J/\psi$ and $\psi^\prime$ are formed, respectively, at 1.2, 6.6, 8.8, 5.8, and 11.0 fm/c after the quark pair are produced in central Au+Au collisions at the top energy of Relativistic Heavy Ion Collider (RHIC), and these times become shorter in semi-central collisions. We further show, as an example, that including the effect of formation time enhances appreciably the survivability of $\Upsilon$(1S) in the produced hot dense matter.
\end{abstract}


\maketitle

\section{introduction}

Studying quarkonium production in relativistic heavy-ion collisions is useful for verifying the existence of a quark-gluon plasma (QGP) in the collision and understanding its properties as first suggested by Matsui and Satz~\cite{Matsui:1986dk}. Studies based on various theoretical models~\cite{Vogt:1999cu,Zhang:2000nc,Zhang:2002ug,Grandchamp:2002wp,Yan:2006ve,Linnyk:2007zx,Zhao:2007hh,Song:2010er,Song:2011xi} have shown that the suppressed production of quarkonia observed in relativistic heavy ion collisions at SPS, RHIC, and LHC~\cite{Adare:2006ns,Abelev:2009qaa,:2010px,Chatrchyan:2011pe,Abelev:2012rv} is mainly due to their dissociation by the thermal partons in the produced quark-gluon plasma, particularly when its temperature is high~\cite{Park:2007zza,Song:2007gm}. Although these studies all give reasonable descriptions of the experimental data, various assumptions and model parameters were introduced. One of these is the time for quarkonia production from initial hard nucleon-nucleon collisions. This time is relevant for determining the survival probability of a quarkonium in the hot dense matter as the heavy quark pair before forming the quarkonium is not likely to be dissociated in a medium either due to the color screening effect or by scattering with the thermal partons as shown in a recent study by one of the authors based on the color evaporation model~\cite{Song:2014qoa}. The survivability of a quarkonium is thus low if its formation time is short, as it is produced in the hot dense matter, and high if its formation time is long, as it is more likely to be formed outside the matter. For quarkonium production from a nucleon-nucleon collision in vacuum, its formation time is known to be not short compared to the time for the production of a heavy quark pair~\cite{Blaizot:1988ec,Karsch:1987zw,Kharzeev:1999bh}. The quarkonium formation time in heavy ion collisions is, however, not well determined. In some studies it is taken to be its value in the vacuum, while in other studies it is assumed to be the same as the thermalization time used in the hydrodynamic approach. It has also been treated simply as a parameter in some studies.

Using the space-time correlator of heavy quark vector currents and the heavy quark potentials extracted from the lattice QCD, we have shown in a recent study that the presence of a QGP makes the quarkonium formation time longer, and this effect becomes stronger with increasing temperature of the QGP~\cite{Song:2013lov}. This study is, however, carried out for a constant QGP temperature, thus neglecting the effect due to the rapid decrease in temperature. In the present study, we extend our previous calculations to include this effect by using the time-dependent QGP temperature obtained from a hydrodynamic model.

This paper is organized as follows: In Sec.~\ref{formulas}, we briefly review the quarkonium formation time in the vacuum and in a QGP. We then describe in Sec.~\ref{results} the quarkonium formation time in a hydrodynamic background for relativistic heavy-ion collisions. Finally, a summary is given in Sec.~\ref{summary}.

\section{formation time of quarkonum in vacuum and in QGP}\label{formulas}

By using the dispersion relation, the space-time correlator of the heavy quark vector current operator can be decomposed into the propagators of physical states as~\cite{Kharzeev:1999bh,Song:2013lov}
\begin{eqnarray}
\Pi(x)\equiv\Pi_\mu^\mu(x)=\frac{3}{\pi}\int dss ~{\rm Im}\Pi(s)D(s,x^2),
\label{correlator}
\end{eqnarray}
where
\begin{eqnarray}
D(s,\tau^2=-x^2)=\frac{\sqrt{s}}{4\pi^2\tau}K_1(\sqrt{s}\tau)
\end{eqnarray}
is the relativistic causal propagator in the coordinate space with $K_1$ being the modified Bessel function and $\tau$ being the Euclidean proper time~\cite{Bogoliubov}. At finite temperature, the factor $\tanh(\sqrt{s}/2T)$ should be multiplied to the right hand side of Eq.~(\ref{correlator}), but its effect is negligible for the present study~\cite{Furnstahl:1989ji,Hatsuda:1992bv,Morita:2007hv}.

Each propagator is weighted by the imaginary part Im$\Pi(s)$ of the heavy quark polarization function, which has contributions from both resonances and the continuum states:
\begin{eqnarray}
{\rm Im}\Pi(s)&=&\sum_i\frac{6e_Q^2}{M_i^2}|\psi_i(0)|^2\frac{\Gamma_i/2}{(\sqrt{s}-M_i)^2+\Gamma_i^2/4}\nonumber\\
&&+\frac{e_Q^2}{4\pi}\theta(\sqrt{s}-\sqrt{s_{\rm th}}).
\label{imaginary}
\end{eqnarray}
In the above, $e_Q$ is the electric charge of a heavy quark in units of $e$; $|\psi_i(0)|$ is the coordinate space wavefunction of the resonance at the origin; and $\sqrt{s_{\rm th}}$ is the threshold energy for the continuum states. The mass distributions of resonances are taken to have the Breit-Wigner form with a peak at $M_i$ and the width $\Gamma_i$, and the continuum states are described by a step function for simplicity.

Substituting Eq.~(\ref{imaginary}) into Eq.~(\ref{correlator}) leads to following contributions from resonances and continuum states to the heavy quark space-time correlator:
\begin{eqnarray}
\Pi_i(\tau)&=&\frac{9 M_i^2}{\pi^2\tau}|\psi_i(0)|^2 K_1(M_i\tau),\nonumber\\
\Pi_{\rm cont}(\tau)&=&\frac{3e^2_Q}{8\pi^4\tau^6}\int^\infty_{\sqrt{s_{\rm th}}\tau}x^4K_1(x)dx.
\label{pi}
\end{eqnarray}
These expressions show that the contribution from the continuum states dominates at early times and is gradually taken over by that from resonances~\cite{Kharzeev:1999bh,Song:2013lov}. The formation time of the ground state is defined as the time when its contribution dominates over all other states. Although $\tau$ is the Euclidean time, the above definition leads to the same formation time or length of a quarkonium as in other approaches~\cite{Kharzeev:1999bh,Gribov:1968gs}.

To calculate the formation time, we first consider the fraction of the ground state in the correlation function, $F(\tau)=\Pi_0(\tau)/\Pi(\tau)$, and then differentiate it with respect to $\tau$, $P(\tau)=dF(\tau)/d\tau$. The function $P(\tau)$ indicates how rapidly the ground state becomes dominant in the correlation function. The formation time is then defined by the expectation value
\begin{eqnarray}
\langle\tau_{\rm form}\rangle=\frac{\int d\tau ~\tau P(\tau)}{\int d\tau ~P(\tau)}.
\label{definition}
\end{eqnarray}

For the formation times of excited states, the contributions from lower-energy states are first subtracted from the correlation function. The fraction of the first excited state, for example, is given by $F(\tau)=\Pi_1(\tau)/(\Pi(\tau)-\Pi_0(\tau))$ such that the contribution from the first excited state, $\Pi_1(\tau)$, eventually dominates~\cite{Kharzeev:1999bh,Song:2013lov}.

According to Eq.~(\ref{pi}), the quarkonium formation time is determined by the three parameters, $M_i$, $|\psi(0)|$, and $\sqrt{s_{\rm th}}$. For charmonium formation in vacuum, its mass is known and the threshold energy can be taken to be twice the D meson or B meson mass. The quarkonium wavefunction at the origin is related to the dielectron decay width of the quarknoium by
\begin{eqnarray}
\Gamma_i^{e^+e^-}=\frac{16\pi\alpha^2 e_Q^2}{M_i^2}~|\psi(0)|^2.
\end{eqnarray}
Using the values determined from the experimental data on $e^+e^-\rightarrow \bar{Q}Q$ for these parameters, the formation times of $J/\psi$ and $\Upsilon$ (1S) in the vacuum are 0.44 and 0.32 fm/$c$, respectively~\cite{Kharzeev:1999bh}.

For quarkonium formation in a QGP, the values of $M_i$, $|\psi(0)|$, and $\sqrt{s_{\rm th}}$ can be determined from solving the Schr\"{o}dinger equation:
\begin{eqnarray}
\bigg[2m_Q-\frac{1}{m_Q}\nabla^2+V(r,T)\bigg]\psi_i(r,T)=M_i\psi_i(r,T),
\end{eqnarray}
using the heavy quark potential $V(r,T)$ and heavy quark mass $m_Q$ in the QGP. According to our recent study based on the QCD sum rules~\cite{Lee:2013dca}, the free energy potential from lattice calculations is the appropriate heavy-quark potential for quarkonium at finite temperature if the quark mass is taken to be its vacuum value. Taking the continuum threshold energy to be $2 m_Q+V(r=\infty, T)$, i.e., the energy of the heavy quark pair when they are infinitely apart, we have evaluated the values of $M_i$, $|\psi(0)|$, and $\sqrt{s_{\rm th}}$ as functions of the QGP temperature.

\begin{figure}[h]
\centerline{
\includegraphics[width=9 cm]{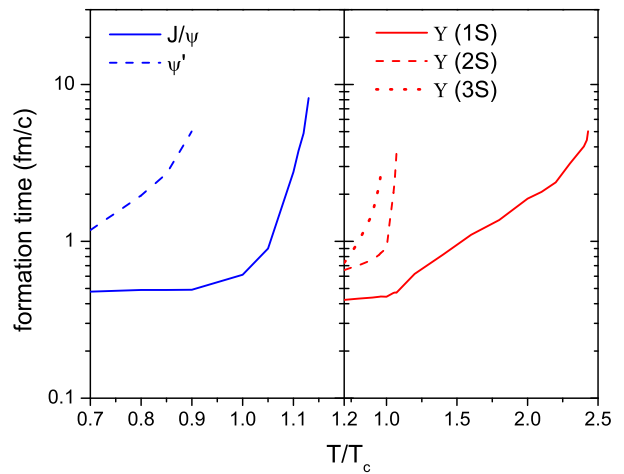}}
\caption{(Color online) Formation times of charmonia (left panel) and bottomonia (right panel) in a QGP as functions of the temperature of the QGP~\cite{Song:2013lov,Song:2014qoa}.}
\label{qgp}
\end{figure}

Figure \ref{qgp} shows the formation times of charmonia and bottomonia in a QGP as functions of its temperature~\cite{Song:2013lov,Song:2014qoa}. We can see that they are longer than their values in vacuum and increase with temperature. These results are consistent with the picture that it takes a longer time for the heavy quark pair to form a quarkoium when its radius becomes larger with increasing temperature~\cite{Song:2013lov}.

\section{quarkonium formation time in relativistic heavy-ion collisions}\label{results}

The above study can be extended to relativistic heavy-ion collisions by taking into account the change of the temperature of the QGP as the quarkonium forms from the heavy quark pair. This is achieved by using the 2+1 ideal hydrodynamic model, assuming the boost invariance and a zero viscosity for the QGP, which we have previously used in studying quarkonia production in relativistic heavy ion collisions~\cite{Song:2011kw,Song:2013tla}. Specifically, we assume that for heavy ion collisions at the top energy of RHIC, hydrodynamics is applicable after an initial thermalization time of about 0.6 fm/$c$~\cite{Song:2011kw,Song:2013tla}. To apply the method described in the previous section to the formation of quarkonia in the nonequilibrium matter before the thermalzation time, we assume for simplicity that the modification of the quarkonium formation time is independent of the momentum anisotropy of the matter~\cite{Epelbaum:2014xea} and is determined by an effective temperature that is related to the entropy density of the matter through the lattice equation of state~\cite{Borsanyi:2010cj} that is used in our hydrodynamic model. Further assuming that the initial nonequilibrium matter expands only in the longitudinal direction, which is supported by studies based on the Color Glass Condensate picture in which the pre-equilibrium state in relativistic heavy-ion collisions is composed of color electromagnetic fields~\cite{McLerran:1993ni,McLerran:1993ka}, the entropy density is then inversely proportional to time.

\begin{figure}[h]
\centerline{
\includegraphics[width=9 cm]{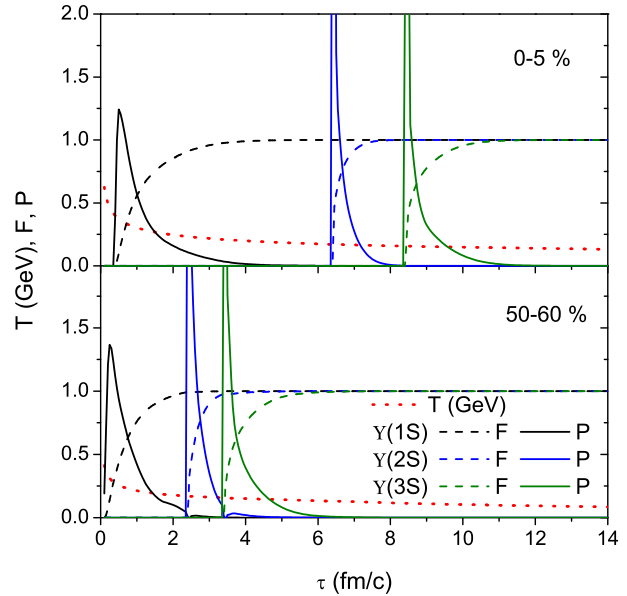}}
\caption{(Color online) Average temperature $T$, the fractions of bottomonium states $F$, and their derivatives $P$ as functions of time in Au+Au collisions at $\sqrt{s_{\rm NN}}=$ 200 GeV for the 0-5 \% (upper panel) and 50-60 \% (lower panel) centralities.}
\label{upsilon}
\end{figure}

Since quarkonium production in heavy ion collisions is from initial nucleon-nucleon binary collisions, we use an average temperature for the hot medium by weighting the local temperature $T(x,y,\tau)$ with the number of binary collisions $N_{\rm binary}(x,y)$ in the transverse plane, that is,
\begin{eqnarray}
T(\tau)=\frac{\int dxdy~ T(x,y,\tau)N_{\rm binary}(x,y)}{\int dxdy~ N_{\rm binary}(x,y)}.
\end{eqnarray}
We calculate $N_{\rm binary}$ using the Glauber model with the nucleon distribution in a nucleus given by a Woods-Saxon function and the empirical nucleon-nucleon scattering cross section as described in Ref.~\cite{Abelev:2008ab}. Shown in Fig.~\ref{upsilon} by the red dotted lines in upper and lower panels are the average temperature of produced matter in Au+Au collisions at $\sqrt{s_{\rm NN}}=$ 200 GeV for the 0-5\% and 50-60\% centralities, respectively. Assuming that the quark pair is produced at $\tau=1/(2m_Q)$, where $m_Q$ is the heavy quark mass and is taken to be 1.5 and 4.75 GeV for charm and bottom respectively, it is seen that the initial temperature in collisions at the 0-5\% centrality is so high that only the continuum states can be formed.

Once the temperature drops to the dissociation temperature of $\Upsilon$(1S) state, the resonance begins to appear near the continuum threshold energy and then completely separates from it. The dashed and solid lines in Fig.~\ref{upsilon} are, respectively, the fraction $F$ of each bottomonium state in the correlator for the bottom quark vector current and its derivative $P$ with respect to $\tau$. The black, blue, and green dashed lines, shown from left to right in the figure, correspond to the 1S, 2S, and 3S bottomonium states, respectively. Each resonance starts to appear at its dissociation temperature, which is 2.5 $T_c$ for the 1S state, 1.1 $T_c$ for the 2S state, and 1.0 $T_c$ for the 3S state, where $T_c=158$ MeV is the phase transition temperature, when the lattice free energy potential is used. In collisions at the 0-5 \% centrality, the system reaches the dissociation temperature at $\tau=0.42$ fm/$c$ for the 1S state, at 6.4 fm/$c$ for the 2S state, and at 8.4 fm/$c$ for the 3S state, and these values change to 0.11 fm/$c$, 2.4 fm/$c$, and 3.4 fm/$c$, respectively, in collisions at the 50-60 centrality. We see that it takes longer for the higher excited state to form. We also see that bottomonia are formed later in central collisions and also take longer than in semi-central collisions. The formation times of the 1S, 2S, and 3S states are, respectively,1.2, 6.6, and 8.8 fm/$c$ in Au+Au collisions at the 0-5\% centrality and 0.7, 2.6, and 3.8 fm/$c$ in collisions at the 50-60 \% centrality. Although the excited states of bottomonium start to form at later times, they complete their formation more quickly than the ground state. This can be understood as follows. Since the heavy quark and antiquark pair produced in a hard collision are initially close in space, it takes time for them to separate to a distance that is comparable to the size of a quarkonium in a medium. It thus take longer for the heavy quark pair to form excited quarkonium states due to their larger sizes. However, these excited states do not start to form before the system reaches their dissociation temperatures when the heavy quark pair are already sufficiently separated, thus requiring less time for them to subsequently form the excited quarkonium states.

\begin{figure}[h]
\centerline{
\includegraphics[width=9 cm]{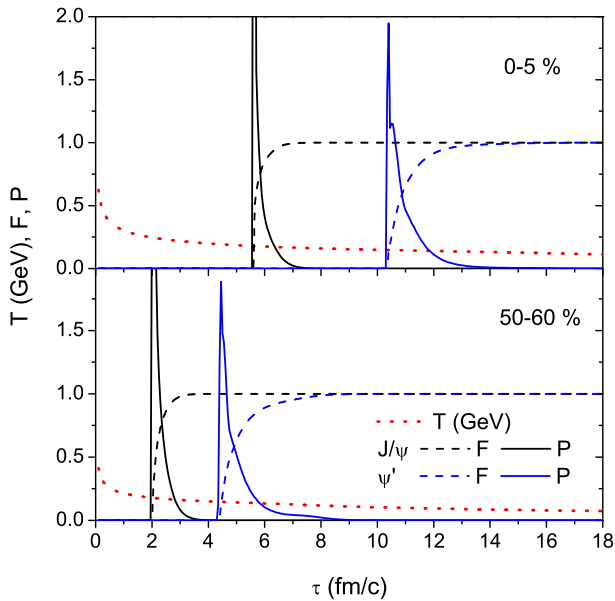}}
\caption{(Color online) Average temperature $T$, the fractions of charmonium states $F$, and their derivatives $P$ as functions of time in Au+Au collisions at $\sqrt{s_{\rm NN}}=$ 200 GeV for the 0-5 \% (upper panel) and 50-60 \% (lower panel) centralities.}
\label{jpsi}
\end{figure}

Figure~\ref{jpsi} shows the charmonia formation time in Au+Au collisions at $\sqrt{s_{\rm NN}}=$ 200 GeV for the 0-5\% and 50-60\% centralities, respectively. The dashed and solid, and dotted lines are, respectively, the fractions $F$ of charmonia, their derivatives $P$, and the average temperature $T$ in collisions corresponding to the two different centralities. Since the dissociation temperature of $J/\psi$ is much lower than that of $\Upsilon$(1S), it appears later at both centralities. With the lattice free energy as the heavy quark potential, the dissociation temperatures of $J/\psi$ and $\psi^\prime$ are about 1.14 and 0.94 $T_c$, respectively. The time for this to happen is 5.6 fm/$c$ for $J/\psi$ and 10.4 fm/$c$ for $\psi^\prime$ in collisions at the 0-5 \% centrality and is 2.0 fm/$c$ for $J/\psi$ and 4.4 fm/$c$ for $\psi^\prime$ in collisions at the 50-60 centrality. Both charmonium states appear after their dissociation temperatures and take, respectively, 5.8 and 11.0 fm/$c$ to form in collisions at the 0-5\% centrality and 2.2 and 5.1 fm/$c$ in collisions at the 50-60\% centrality.

Although the initial energy density in heavy-ion collisions is so high that the heavy quark and antiquark pair produced in a hard collision are screened from each other, there is still a possibility for them to form a bound state if the duration of the high density stage is short, because they are initially very close to each other in space. However, if the duration of high energy density is long, it is then unlikely that the initial heavy quark and antiquark pair can form a quarkonium as they are far apart at the formation time, such as for the excited states of bottomonia and all charmonia in central collisions as well as the excited states of both bottomonia and charmonia in semi-central collisions. In this case, quarkonia production in heavy ion collisions will be dominated by the regeneration from thermalized heavy quarks and antiquarks in the QGP.

\begin{figure}[h]
\centerline{
\includegraphics[width=9 cm]{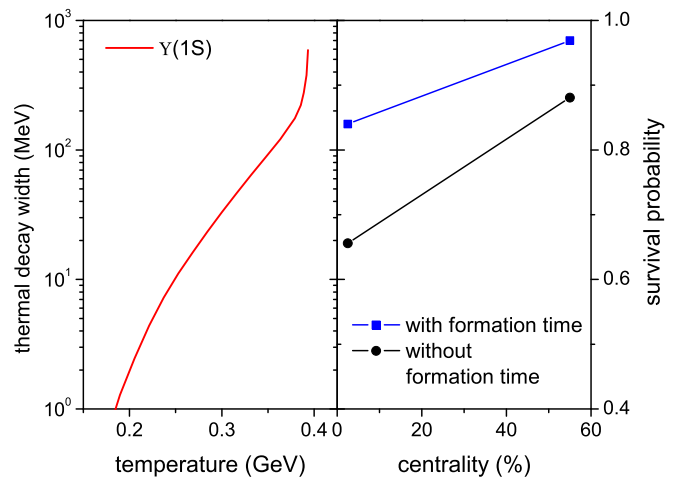}}
\caption{(Color online) Thermal decay width of $\Upsilon$ (1S) as a function of temperature (left panel) and its survival probabilities from thermal decay with and without including the formation time as functions of collision centrality (right panel).}
\label{survival}
\end{figure}

After its formation, a quarkonium can be dissociated by thermal partons. Its survivability in a heavy ion collision is given by
\begin{eqnarray}
S=\frac{\int_0^\infty d\tau P(\tau)\exp{\left(-\int_\tau^{\tau_c}\Gamma(\tau^\prime)d\tau^\prime\right)}}{\int_0^\infty d\tau P(\tau)},
\end{eqnarray}
if we neglect the thermal dissociation below $T_c$ as its effect is much smaller than that above $T_c$. In the above, $P(\tau)$ is the fraction of quarkonium formed per unit time at $\tau$ as defined above Eq.(\ref{definition}) and $\tau_c$ is the time when the quark-gluon plasma stage ends. The dissociation width $\Gamma(\tau)$ of the formed quarkonium at time $\tau$ in the above equation depends on the temperature of the quark-gluon plasma at that time and its dissociation cross section by thermal partons. For the latter, it can be calculated from the pQCD with massive thermal partons as in Ref.~\cite{Park:2007zza} and used extensively in Refs.~\cite{Song:2010ix,Song:2010er,Song:2011xi,Song:2011nu,Song:2013tla}. As an example, we show in the left panel of Fig.~\ref{survival} the dissociation width of $\Upsilon$(1S) as a function of temperature calculated with the lattice free energy as the potential between the bottom quark and antiquark pair in the quark-gluon plasma. In the right panel of Fig.~\ref{survival}, we show by filled squares the survival probabilities of $\Upsilon$(1S) in Au+Au collisions at $\sqrt{s_{\rm NN}}=200$ GeV for both centralities of 0-5\% and 50-60\%. Compared to those for $\Upsilon$(1S) of zero formation time (filled circles), i.e., the $\Upsilon$ (1S) is formed as soon as the temperature drops to its dissociation temperature, including the formation time increases appreciably the survivability of $\Upsilon$(1S) in relativistic heavy ion collisions, particularly in central collisions.

\section{summary}\label{summary}

The formation time of quarkonium is one of essential ingredients used in studies on quarkonium production in relativistic heavy-ion collisions. To calculate the formation time in the quark-gluon plasma produced in relativistic heavy-ion collisions, we have extended the method which uses the space-time correlator for the heavy quark vector current and its dispersion relation to the case that the produced quark-gluon plasma is described by a 2+1 ideal hydrodynamic model. Before the initial thermalization time where hydrodynamics is not applicable, we have assumed that the hot dense matter expands only in the longitudinal direction and its effect on quarkonium formation can be described by an effective temperature. We have found that the quarkonium states begin to appear as the hot dense matter expands and cools to their dissociation temperatures, and the times to complete their formations in central Au+Au collisions at the top energy of RHIC are at 1.2, 6.6, 8.8, 5.8, and 11.0 fm/c for $\Upsilon$(1S), $\Upsilon$(2S), $\Upsilon$(3S), $J/\psi$ and $\psi^\prime$, respectively. In semi-central or peripheral collisions, both the start time of the formation and the formation time of a quarkonium are shorter than in central collisions.

The increased formation time of quarkonia in the hot dense matter is expected to affect their survivability in relativistic heavy ion collisions. We have found, for example, that including the formation time increases appreciably the survivability of $\Upsilon$(1S). This effect depends, however, on the magnitude and temperature dependence of the thermal decay width of a quarkonium, with a strong temperature dependence giving a larger effect and a weak temperature dependence giving a small effect. Unfortunately, both the magnitude and temperature dependence of the quarkonium thermal decay widths are presently not well determined as very different values have been used in the phenomenological studies of qaurkonium production in relativistic heavy ion collisions.

Finally, since the formation time of $\Upsilon(1S)$ in relativistic heavy ion collisions is not much longer than the initial thermalization time, a better treatment of the effect of the initial nonequilibrium matter than that used in the present study is needed.

\section*{Acknowledgements}

This work was supported in part by the U.S. National Science Foundation under Grant No. PHY-1068572, the US Department of Energy under Contract No. DE-FG02-10ER41682 within the frame of the JET Collaboration, the Welch Foundation under Grant No. A-1358, the Korean Research Foundation under Grant Nos. KRF-2011-0020333 and KRF-2011-0030621, and the DFG.

\end{document}